\documentclass[prd,amsfonts,twocolumn,nofootinbib,showpacs]{revtex4}
\RequirePackage{graphicx,  amsmath}
\begin{document}
\title{Free light fields can change the predictions of hybrid inflation}
\author{Tomohiro Matsuda}
\affiliation{Department of Physics, Lancaster University,  
Lancaster LA1 4YB, UK, and
 Laboratory of Physics, Saitama Institute of Technology,
Fukaya, Saitama 369-0293, Japan}

\begin{abstract}
\hspace*{\parindent}
We show that the free light scalar fields that may exist in the
 inflationary Universe
can change the predictions of the hybrid inflation model.
Possible signatures are discussed, which can be used to discriminate the
 sources of the spectrum. 
\end{abstract}
\maketitle
\section{Introduction}
Inflation generates the source of the large-scale perturbation that is
needed to explain the inhomogeneity and the structure of the
Universe \cite{Lyth-book}. 
In the original scenario of the inflationary Universe,
 the adiabatic mode of the
inflaton fluctuation sources the curvature perturbation when the
perturbation leaves the horizon. 

However, if there are many light scalar fields during inflation, inflation
may create isocurvature perturbations for these fields, and those
isocurvature perturbations may
cause significant creation of the curvature perturbation after
the horizon exit.

The generation of the curvature perturbation at the end of
inflation \cite{At-the-end} is basically caused by (1) modulation of the
coupling constants or (2) two-field hybrid inflation in which both 
inflaton fields are coupled to the waterfall field.
One thing that is common in the previous models of the ``end of
inflation'' scenario is that {\bf the extra light field has an explicit 
 coupling to the inflaton sector} \cite{multi-brid}.
One might think that this requirement is essential and quite obvious;
however {\em we are considering in this paper the removal of this
basic requirement.}
In our model the additional fields are not coupled to the inflaton
sector, but it 
changes the definition of the adiabatic field.
As a result, the inflation end is not identical to the uniform
density hypersurfaces.
 
First consider a model in which the inflaton $\phi$ and the waterfall field
$\chi$ have the hybrid-type potential given by
\begin{equation}
\label{base-pot}
V(\phi,\sigma)=\frac{\lambda^2}{4}\left(\chi^2-M^2\right)^2
+\frac{g^2}{2}\chi^2 \phi^2 +\frac{1}{2}m_\phi^2 \phi^2.
\end{equation}
This defines the ``inflaton sector'' of the model.
Suppose that inflation starts with $\phi>\phi_c$, and the end of
inflation is defined by $\phi=\phi_c$, where the waterfall begins.
The critical point $\phi_c$ is given by
\begin{equation}
\phi_c\equiv \frac{\lambda}{g}M.
\end{equation}
The number of e-foldings is given by
\begin{equation}
N=H\int_{\phi_*}^{\phi_e}\frac{d\phi}{\dot{\phi}}\simeq\frac{1}{\eta}\ln
\frac{\phi_*}{\phi_c}.
\end{equation}
Here $\eta$ is the slow-roll parameter defined by
\begin{equation}
\eta\equiv \frac{V_{\phi\phi}}{3H^2},
\end{equation}
where $H$ is the Hubble parameter during inflation and the subscript
means the derivative with respect to the field.
In the single-field inflation model we always find the trivial
coincidence between the uniform density hypersurfaces and the end of
inflation.
In that case, one cannot find the
perturbation $\delta N_e$  created at the
end~\cite{matsuda-warm,matsuda-gra,infla-curv}.
Here $\delta N_e$ measures the discrepancy between
the uniform density hypersurfaces and the end of inflation.

Going back to the past model~\cite{At-the-end}, generation of the
perturbation at the end of inflation is considered 
for a two-field model.
In that case the additional light field is coupled to the
waterfall field and that causes $\delta N_e\ne 0$ at the end.
Namely, if one introduces another inflaton $\phi_2$ that has the
same interaction as the primary field $\phi_1$, the end of inflation 
depends on $\phi_2$ as;
\begin{equation}
\phi_1^2+\phi_2^2=\phi_c^2.
\end{equation}
Then, suppose that $\phi_2$ is much lighter than $\phi_1$, the entropy
perturbation $\delta s\simeq \delta \phi_2\ne 0$ creates the
perturbation $\delta \phi_1 = -\frac{\phi_2}{\phi_c}\delta \phi_2$ at
the end.
This leads to the perturbation of the number of e-foldings at the end of
inflation, which is given by
\begin{equation}
\delta N_e \equiv -\left[\frac{1}{\eta}\frac{\delta \phi_1}{\phi_1}
\right]_e
\ne 0.
\end{equation}
This is the usual scenario of $\delta N_e\ne 0$.

In this paper, we consider a similar mechanism of generating curvature
perturbation at the end of inflation, {\em but in contrast to the
usual scenario, the additional fields are decoupled from the inflaton sector}.

\begin{figure}[t]
\centering
\includegraphics[width=0.8\columnwidth]{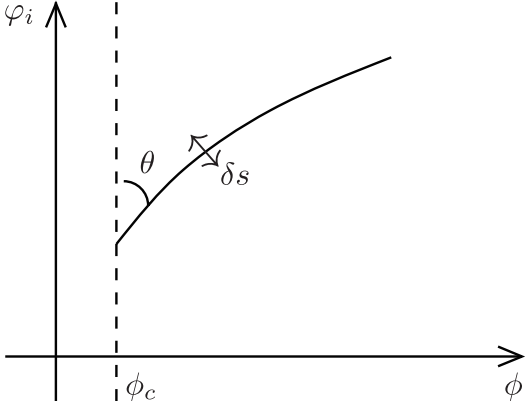}
 \caption{Free light scalar fields ($\varphi_i$) are added 
to the conventional hybrid-type potential. $\delta s$ is the entropy
  perturbation, and $\theta$ measures
  the velocity ratio between $\phi$ and the others.
Note that $\theta=\pi/2$ is assumed in the original hybrid inflation, but
the assumption is not quite obvious when there are many light scalar
 fields in the model, even if they are decoupled from the hybrid-type
 potential.} 
\label{fig:plusone}
\end{figure}

Keeping the inflaton $\phi$ in the potential given by
Eq.(\ref{base-pot}),
we introduce {\bf free} light fields ($\varphi_i$), which have
no explicit interaction with the waterfall field $\chi$.
Although these fields ($\varphi_i$) are not coupled to the ``inflaton
sector'', they must participate in the definition of the
adiabatic inflaton field;  
$\dot{\sigma}^2 \equiv \dot{\phi}^2+\sum_{i=1}^{N_f}\dot{\varphi}^2_i$,
which is mandatory.
A typical situation is shown in Fig.\ref{fig:plusone}.

In this paper we consider the perturbation caused by the entropy
perturbation $\delta s$.
The angled trajectory ($\theta\ne \pi/2$) causes significant creation of
the curvature perturbation at the end.

\section{Free light scalars in hybrid inflation}
One might be skeptical about the generation of the curvature
perturbation at the end, if it is realized just by adding free scalar
fields to the model. 
To illustrate what happens in this model, we show the simplest calculation, in
which one ``standard'' inflaton $\phi$ and one additional
light field $\varphi$ have the same quadratic potential with the same
mass $m_\phi=m_\varphi \equiv m$. 
Therefore, the potential is given by
\begin{equation}
V(\phi,\sigma)=\frac{\lambda^2}{4}\left(\chi^2-M^2\right)^2
+\frac{g^2}{2}\chi^2 \phi^2 +\frac{1}{2}m^2 (\phi^2+\varphi^2).
\end{equation}

\begin{figure}[b]
\label{fig2nd}
\centering
\includegraphics[width=0.8\columnwidth]{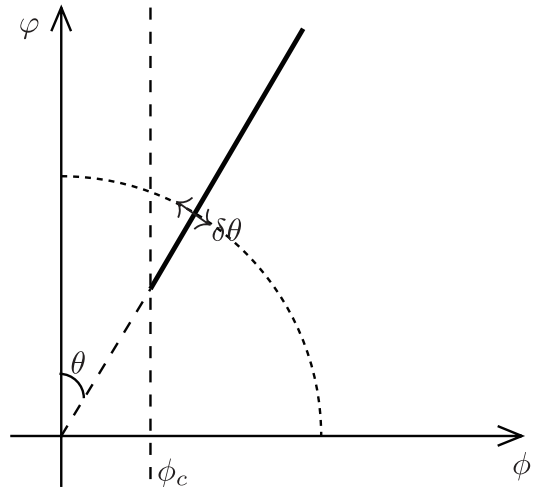}
 \caption{The entropy perturbation leads to $\delta \theta$, which
  appears along the uniform density hypersurfaces (dotted circle).}
\label{fig:angle}
\end{figure}

This assumption makes the trajectory straight, and removes ambiguities
related to the possible non-trivial evolution of the perturbations 
during inflation. 
(See Fig.\ref{fig2nd}.)
Defining the adiabatic field $\sigma^2 \equiv \phi^2 + \varphi^2$,
we find
\begin{eqnarray}
\phi&=&\sigma \sin \theta\\
\varphi &=& \sigma \cos\theta.
\end{eqnarray}

In this model, the end of inflation defined by the constant
$\phi=\phi_c$ does not coincide with the uniform density hypersurfaces 
defined by constant $\sigma$.
We find the end is given by
\begin{equation}
\sigma_e(\theta)\equiv \frac{\phi_c}{\sin\theta},
\end{equation}
which is perturbed by $\delta \theta\ne 0$.
Therefore, although $\phi_c$ is not perturbed in this model, the entropy
perturbation ($\delta \theta \ne 0$) 
causes $\delta N_e\ne 0$ at the end of inflation.

The inhomogeneous end of inflation
caused by the entropy perturbation is thus given by
\begin{eqnarray}
\delta \sigma_e &\equiv&
\frac{\partial \sigma}{\partial \theta}\delta \theta+
\frac{1}{2}\frac{\partial^2 \sigma}{\partial \theta^2}(\delta
\theta)^2+...\nonumber\\
&=&-\left[
\frac{\cos\theta}{\sin^2\theta}
\right]\phi_c\delta \theta 
+\frac{1}{2}\left[
\frac{1+\cos^2\theta}{\sin^3 \theta}
\right]
\phi_c\left(\delta
	     \theta\right)^2+..,\nonumber\\
&\simeq& -\frac{\cos\theta \phi_c}{\sin^2
		   \theta}\left(\frac{\delta s}{\sigma}\right)_*
+\frac{1}{2}\frac{(1+\cos^2\theta)\phi_c}{\sin^3 \theta}\left(\frac{\delta s}{\sigma}
\right)_*^2\nonumber\\
&=& -\frac{1}{\tan\theta}\frac{\sigma_e}{\sigma_*}\delta s_*
+\frac{1+\cos^2\theta}{2\sin^2 \theta}\frac{\sigma_e}{\sigma_*^2}
\left(\delta s_*\right)^2,
\end{eqnarray}
where the subscripts ``$e$'' and ``$*$'' denote the value at the end of
inflation 
and at the epoch when the perturbation leaves the horizon, respectively.
In this simplest example, the relation $\delta \theta =
\left[\frac{\delta s}{\sigma}\right]_*$ is exact and $\delta \theta$ is
constant after the horizon exit.
Note that in this formalism $\delta 
s_*$ is Gaussian but $\delta \sigma_e$ is not always a Gaussian
perturbation.

The curvature perturbation generated at the end of inflation is thus given by
\begin{eqnarray}
\delta N_e&\equiv& H\frac{\delta \sigma_e}{\dot{\sigma}_e}
= -\left[\frac{1}{\eta}\frac{\delta \sigma}{\sigma}\right]_e\nonumber\\
&\simeq&
\frac{1}{\eta_e}\frac{1}{\tan\theta}
\left(\frac{\delta s}{\sigma}\right)_*
-\frac{1+\cos^2\theta}{2\eta_e\sin^2 \theta}
\left(\frac{\delta s}{\sigma}
\right)_*^2.
\end{eqnarray}
Since the perturbation of the inflation generated at the
horizon exit is given by
\begin{equation}
\delta N_* = \left[\frac{1}{\eta} \frac{\delta \sigma}{\sigma}\right]_*,
\end{equation}
we find the ratio between the ``initial'' and ``at the end'' perturbations;
\begin{eqnarray}
r&\equiv& \left|\frac{\delta N_e}{\delta N_*}\right|
=\left|\frac{\dot{\sigma}_*}{\dot{\sigma}_e}\frac{\delta
  \sigma_e}{\delta \sigma_*}\right| 
\nonumber\\
&\simeq& \frac{\eta_*}{\eta_e}\frac{1}{\tan\theta}
= \frac{1}{\tan\theta},
\end{eqnarray}
where the ratio is for the first order perturbations and
 $\delta s_*=\delta \sigma_*$ is used for the calculation.
We can see that $\delta N_e$ dominates (i.e, $r>1$ is realized) when
$\tan\theta <1$.

The spectral index $n-1\equiv \partial{\cal P}_\zeta/\partial \ln k$ for
the perturbation $\delta N_e$ is~\cite{At-the-end}
\begin{equation}
\label{index-end}
n-1\simeq -2\epsilon_H+2\eta_s,
\end{equation}
where $\eta_s\simeq \eta_{\sigma}\equiv V_{\sigma\sigma}/3H_*^2$ and 
$\epsilon_H\equiv \dot{H}_*/H^2_*$ are defined at the horizon exit.
{\em The spectral index is obviously different from the
standard hybrid inflation scenario.}

Using the second order perturbation, we can estimate the non-Gaussianity 
parameter
\begin{equation}
f_{NL}\sim \frac{\eta_e}{\cos^2\theta},
\end{equation}
where $\theta\ll 1$ and $\eta_e \ll 1$ suggest $f_{NL}\ll 1$.
Here the definition of $f_{NL}$ is
\begin{equation}
f_{NL} \simeq \frac{5}{6} \frac{N_i N_j N_{ij}}{(N_mN_m)2}
+ \cdots
, \end{equation}
where a subscript $i$ denotes the derivative with respect to the $i$-th
field, and ``$\cdots$'' includes loop corrections that are usually negligible.

Since $f_{NL}$ is obviously small for the quadratic potential,
in which $\eta_*=\eta_e$ is mandatory, 
we need to consider non-quadratic potential for the enhancement.

One way to enhance $f_{NL}$ is to consider
the effective potential that is dominated by the higher polynomial;
\begin{equation}
V(\phi,\varphi)\simeq M^4+\alpha_n\sigma^p.
\end{equation}
This potential may allow $\eta_*\gg \eta_e$, which makes it possible to find
$r>1$ and $f_{NL}>1$ at the same time.

\subsection{More fields}
If there are $N_f$ light scalar fields ($\varphi_i$, $i=1,...N_f$) in the inflationary
Universe, a naive statistical 
expectation is
$\theta\sim 1/N_f\ll 1$.
To avoid the significant creation of $\delta N_e$ at the end,
these scalar fields must be settled in their minima before the end of
inflation, or should have
very flat potential compared 
with the inflaton, so that  $\sum_{i=1}^{N_f}\dot{\varphi}_i^2\simeq 0$
 is a plausible approximation during inflation.

If there are too many fields in the Universe,
N-flation \cite{N-flation-paper} may start before the onset of hybrid inflation. 
Then hybrid inflation may begin when $M^4$ starts dominating the
Universe.

\section{Free light fields in the spectrum}
\begin{figure}[t]
\centering
\includegraphics[width=0.3\columnwidth]{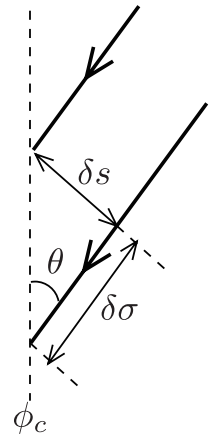}
 \caption{Closer look at the intersection in Fig.\ref{fig:plusone}. 
The perturbed trajectory is approximately parallel when $\delta
 \theta\ll \theta$.} 
\label{fig:general}
\end{figure}

Above we considered the simplest set-up for the model,
in which only one field (adiabatic field) appears in the effective
potential during inflation, while the end is caused by the coupling
between $\phi$ and $\chi$.
However, in more general cases there may be many fields that
may have positive/negative mass terms or higher polynomials.
Even in that case, the first order
perturbation requires very simple parameters.
From Fig.\ref{fig:general}, we can see that for $\delta \theta\ll
\theta$ the perturbation at the end of inflation is given by
\begin{equation}
\delta \sigma_e \simeq \frac{\delta s_e}{\tan\theta},
\end{equation}
where we assume $\delta s_e\sim \delta s_*$.
Here $\theta$ is defined by
\begin{equation}
\tan\theta \equiv
 \frac{|\dot{\phi}|}{\sqrt{\sum_{i=1}^{N_f}\dot{\varphi}_i^2}},
\end{equation}
and the adiabatic field $\sigma$ is 
\begin{equation}
\dot{\sigma}^2\equiv \dot{\phi}^2+\sum_{i=1}^{N_f}\dot{\varphi}_i^2.
\end{equation}
More general definitions of the adiabatic field (probably including
non-canonical kinetic terms \cite{matsuda-non-cano})
would be interesting, but they are not the topic
in this paper.
The curvature perturbation generated at the end of inflation is thus
given by
\begin{eqnarray}
\delta N_e &\simeq& H \frac{\delta \sigma_e}{\dot{\sigma}_e}\nonumber\\
&\simeq& \left[\frac{1}{\sqrt{8\pi^2\epsilon_{e}}}\frac{H}{M_p}\right]\times
 \frac{1}{\tan\theta_e}.
\end{eqnarray}
Comparing the result with the curvature perturbation generated at the
horizon exit ($\delta N_*$), we find that 
$\delta N_e$ is enhanced by the factor $1/\tan\theta_e\gg 1$.

If one compares this result with the ``standard'' calculation
\begin{equation}
\delta N_0\equiv \left[H\frac{\delta \phi}{\dot{\phi}}\right]_*
\simeq  \left[\frac{1}{\sqrt{8\pi^2\epsilon_\phi}}\frac{H}{M_p}\right],
\end{equation}
where the slow-roll parameter $\epsilon_\phi$ is defined by
$\epsilon_\phi\equiv \frac{1}{2}M_p^2 \left(\frac{V_\phi}{V}\right)^2$,
one finds 
\begin{eqnarray}
\frac{\delta N_e}{\delta N_0}&\simeq &
\sqrt{\frac{\epsilon_\phi}{\epsilon_e}}
\frac{1}{\tan\theta_e}.
\end{eqnarray}
For the simplest (two field, equal mass) model we find
\begin{equation}
\sqrt{\frac{\epsilon_\phi}{\epsilon_e}}\simeq \sin\theta,
\end{equation}
which leads to
\begin{eqnarray}
\frac{\delta N_e}{\delta N_0}&\simeq &
\cos\theta<1,
\end{eqnarray}
or 
\begin{eqnarray}
\frac{\delta N_*}{\delta N_0}&\simeq &
\sin\theta<1.
\end{eqnarray}
The minimum ratio $\frac{\delta N_e}{\delta N_0}\simeq\frac{\delta
N_*}{\delta N_0}\simeq1/\sqrt{2}$ is obtained when $\tan\theta=1$, where
both perturbations contribute (i.e, $\delta N_*\simeq \delta N_e$).
Obviously, the additional light field changes the spectrum.

\section{Conclusions and discussions}
In the inflationary Universe we may expect many light scalar fields
$\varphi_i$.
Since the adiabatic field during inflation must be defined using all the
fields that are moving  
during inflation, the mismatch between the end of hybrid inflation
(usually defined by 
$\phi$ that is coupled to the waterfall field)
and the uniform density hypersurfaces (always defined by the adiabatic field)
causes creation of $\delta N$ at the end of inflation.
Using the conventional $\delta N$ formalism, we showed that the fields
that are decoupled from the inflaton sector can play
significant role in creating the curvature perturbation.

The signs of the light fields may appear in the spectral
index and/or in the non-Gaussianity, which can be used to discriminate
the origin of the perturbation.
The precise calculation of these parameters is quite difficult when
there are many fields during 
inflation, but our modest prediction is that the deviation from the
standard prediction may indicate 
the presence of light scalar fields in the early Universe.

In this paper hybrid inflation is simplified assuming that 
it ends with an instantaneous waterfall starting at the critical
instability point $\phi_c$. 
This simplification is not valid if inflation can continue for more
than 60 e-folds during the waterfall~\cite{hyb-alt}.
However, the model that spends more than 60 e-folds during the waterfall
should correspond to hilltop inflation (with probably some
modulation caused by the interaction with
$\phi$~\cite{modulated-inflation}) that should be 
discriminated from the conventional hybrid inflation scenario.

We discussed the physics related to the evolution before the waterfall
using the simplification of the instantaneous waterfall.
Our simplification is valid only when the end of inflation
mechanism~\cite{At-the-end} is valid for the hybrid inflation model.

\section{Acknowledgment}
We wish to thank K.Shima for encouragement, and our colleagues at
Nagoya university and Lancaster university for their kind hospitality.

\end{document}